\documentclass[epj]{svjour}
\usepackage{graphics}
\usepackage[usenames,dvipsnames]{color}
\usepackage{epsf}
\usepackage{bm}
\usepackage{dcolumn}
\usepackage{latexsym}
\usepackage{amsmath}
\usepackage{amsfonts}
\usepackage{amssymb}
\usepackage{graphicx}
\usepackage[active]{srcltx}

\usepackage{bm}
\def\iems#1{}

\def\ins#1{{\it #1}}

\def\meq#1{}
\def\meq#1{}

\def\ins#1{}

\def\comment#1{}

\def\mn#1{\marginpar[]{\scriptsize#1}}
\def\mn#1{}

\usepackage{ulem}
\usepackage{cancel}
\usepackage{color}
\def\rr#1{\textcolor{red}{#1}}

\def\mn#1{\marginpar[\tiny{\rr{#1}}]{\tiny{\rr{#1}}}}

\def\rr#1{}

\usepackage{hyperref}
\newcommand{\be}{\begin{equation}}\newcommand{\ee}{\end{equation}}
\newcommand{\bea}{\begin{eqnarray}}\newcommand{\eea}{\end{eqnarray}}
\newcommand{\beaa}{\begin{eqnarray}}\newcommand{\eeaa}{\end{eqnarray}}
\newcommand{\ba}{\begin{array}}\newcommand{\ea}{\end{array}}
\newcommand{\bit}{\begin{itemize}}\newcommand{\eit}{\end{itemize}}
\newcommand{\ben}{\begin{enumerate}}\newcommand{\een}{\end{enumerate}}
\def\be{\begin{equation}}
\def\ee{\end{equation}}
\def\bea{\begin{eqnarray}}
\def\eea{\end{eqnarray}}
\def\bear{\begin{array}}
\def\eear{\end{array}}

\def\a{\alpha}

\def\b{\beta}
\def\d{\delta}

\def\x5{x^{5}}

\def\C{C}
\def\S{S}
\def\a{\mu}
\def\b{\nu}
\def\c{\lambda}
\def\d{\kappa}

\definecolor{darkred}{rgb}{.8,0,0}

\definecolor{darkblue}{rgb}{0,0,.7}

\begin{document}
\title{Inflationary cosmology from quantum  Conformal Gravity}
\author{Petr Jizba\inst{1,}\inst{2,}\thanks{e-mail: p.jizba@fjfi.cvut.cz},
Hagen Kleinert\inst{2,}\thanks{e-mail: h.k@fu-berlin.de}
\and Fabio Scardigli\inst{3,}\inst{4,}
\thanks{e-mail: fabio@phys.ntu.edu.tw}%
}                     
%
%
\institute{FNSPE, Czech Technical
University in Prague, B\v{r}ehov\'{a} 7, 115 19 Praha 1, Czech Republic
\and ITP, Freie Universit\"{a}t Berlin, Arnimallee 14, D-14195 Berlin, Germany
\and Department of Mathematics, College of Engineering, American University of the Middle East,\\
P.O.Box 220 Dasman, 15453 Kuwait
\and Dipartimento di Matematica, Politecnico di Milano, Piazza Leonardo da Vinci 32, 20133 Milano, Italy}
\date{Received: date / Revised version: date}
%
\abstract{
We analyze the functional integral for  quantum Conformal Gravity and show that
with the help of a Hubbard--Stratonovich transformation, the action can be broken into a local quadratic-curvature theory coupled to a scalar field. A one-loop effective action calculation reveals that strong fluctuations of the metric field are capable of spontaneously generating  a dimensionally transmuted parameter which, in the weak-field sector of the  broken phase, induces a Starobinsky-type $f(R)$-model with a gravi-cosmological constant. A resulting non-trivial relation between Starobinsky's
parameter and the gravi-cosmological constant is highlighted and implications for cosmic inflation are briefly discussed and compared with the recent PLANCK and BICEP2 data.
} 
\maketitle
%
\section{Introduction}
\label{intro}
%
The idea that Einsten's gravity may be considered as
a large-distance effective theory arising from a spontaneous or dynamical
symmetry breakdown in some underlying scale
invariant quantum field theory dates back to works of
Minkowski~\cite{Minkowski}, Smolin~\cite{Smolin},  Adler~\cite{Adler}, Zee~\cite{Zee},
Spokoiny~\cite{Spokoiny}, Kleinert and Schmidt~\cite{SK}, and others
(see, e.g., Ref.~\cite{Duff} for recent review), even though
the motivations can be traced back to 1960's  seminal papers of Zeldovich~\cite{Zeldovich}
and Sakharov \cite{Sakharov}.
The ensuing mechanisms for symmetry breaking
are realized typically by spontaneously breaking a scale invariance
in appropriate scale-invariant quantum field theory propagating in a curved
spacetime~\cite{Adler}
or by a Conformal Gravity
(CG) which is dynamically broken via additional scalar
fields~\cite{Antoniadis,Modesto}.

In particular, CG has recently attracted renewed attention
because local conformal invariance seems to be the key component
in a number of
cosmological models.
This activity was substantially
fueled by Mannheim {\it et al.}
no-ghosts result~\cite{B-M,B-MII,Tkach,note2,note2b}, Smilga's benign-ghost result~\cite{Smilga},
new non-perturbative approaches~\cite{Ambjorn:12}
and by related works on conformal anomaly~\cite{Mann:12}.
The CG has been since revisited from various points of
view, e.g., as an alternative to standard Einstein gravity
giving a (partial) resolution of a flatness problem~\cite{Mann:92},
or as an explanatory frame for missing matter in galaxies~\cite{Brian:12}
and a possibly vanishing cosmological constant~\cite{MANN}.
The CG has also been explored recently in a number of
theoretical and observational frameworks including
conformal supergravity~\cite{Tseytlin}, Twistor-String theory~\cite{witten}, asymptotic safety
theories~\cite{Weinberg,Weiberg2}, black-hole complementarity issue~\cite{THOOFT}, \
AdS/CFT correspondence~\cite{Hyun}, and the type Ia
supernova (SNIa) and $H(z)$ observational data~\cite{Yang:13}.

Unfortunately, the particle-spectrum of CG does not contain (at least not on-shell) a scalar field.
In fact, CG has 6 (on-shell) propagating degrees of freedom; massless spin-2 graviton,
massless spin-1 vector boson and massless spin-2 ghost field~\cite{Tseytlin,Riegert}.
Should the Einstein gravity be induced within CG at low energies,
the absence of a fundamental scalar poses immediately two problems: a)
it is difficult to break a conformal symmetry (either spontaneously or dynamically) without
a fundamental spinless boson~\cite{Tseytlin}, b) scalar degree of freedom is of a central
importance to generate correct primordial density perturbations during inflation~\cite{Duff}.
For these reasons an external scalar field is sometimes artificially coupled to CG~\cite{Antoniadis,Modesto}. In this paper we wish to point out a subtle fact that a non-dynamical spurion scalar field can be introduced in CG via the
Hubbard--Stratonovich transformation  without spoiling particle spectrum, (non-perturbative) unitarity, and renormalizability of CG. The spurion field is actually an imprint of
a scalar degree of freedom that would normally be present in the theory if the (local) conformal symmetry
would not decouple it from the on-shell spectrum. The spurion field morphs into a physical scalar field
(scalaron or gravi-scalar) after
its kinetic term gets generated radiatively. The field then mediates a dynamical breakdown of the conformal symmetry.
%
In the broken phase the scalaron field acquires a
non-trivial vacuum expectation value (VEV) via dimensional transmutation.
The resulting low-energy behavior in the broken phase can be identified with Starobinsky's
$f(R)$-model (SM) with a gravi-cosmological constant or, in a dual picture, with a two-field hybrid inflationary model.
A scalaron field helps to form (composite) inflaton, and assists during the inflaton decay in the reheating phase.
%
%


\section{Quantum Conformal Gravity}
\label{sec:1}
CG is a pure metric theory that possesses general coordinate invariance,
which augments standard gravity with the additional Weyl symmetry, i.e.,
invariance under a local rescaling  of the metric $g_{\mu\nu}(x)\rightarrow e^{2\alpha(x)}g_{\mu\nu}(x)$,
with  $\alpha(x)$ being an arbitrary local function.
The simplest CG action, i.e., action with both reparametrization and Weyl invariance reads~\cite{Weyl1,Bach1}

\begin{equation}
{\cal A}_{\rm conf} \ = \ -\frac1 {8\alpha_c^2}\int d^4x\, (-g)^{1/2}C_{\lambda\mu\nu\kappa}
C^{\lambda\mu\nu\kappa}\, .
\label{PA1}
\end{equation}
\vspace{-0.3mm}
Here $\alpha_c$ is a dimensionless coupling constant (in natural units) and $C_{\lambda\mu\nu\kappa}$ is the
{\it Weyl tensor\/} which in $4$ space-time dimensions reads
\begin{eqnarray}
C_{\lambda\mu\nu\kappa} \ &=& \
R_{\lambda\mu\nu\kappa} \ - \ \left(g_{\nu[\lambda}R_{\mu]\kappa} -
g_{\kappa[\lambda}R_{\mu]\nu} \right)
\nonumber\\[1mm]
&+& \frac{1}{3}R \ \! g_{\nu[\lambda}g_{\mu]\kappa}\, ,
\label{P2a}
\end{eqnarray}
with $R_{\lambda\mu\nu\kappa}$ being the {Riemann curvature tensor},
$R_{\lambda\nu}=R_{\lambda\mu\nu}{}^\mu$  the  {Ricci tensor}, and $R\equiv R_\mu{}^\mu$
the {scalar curvature}. Throughout we adopt signature $(+,-,-,-)$ and sign conventions of Landau--Lifshitz.
With the help of the Gauss--Bonnet  theorem one
can cast ${\cal A}_{\rm conf}$ into equivalent form (modulo topological term)
\begin{equation}\!
{\cal A}_{\rm conf} \ = \
-\frac1{4\alpha_c^2}\int d^4x\, (-g)^{1/2}\left[R_{\mu\kappa}R^{\mu\kappa}-
\frac{1}{3} R^2\right].
\label{PA2}
\end{equation}
%
%

Variation of ${\cal A}_{\rm conf}$ with respect to the metric
yields Bach's field equation~\cite{Bach1}
\begin{equation}
2D_{\lambda} D_{\kappa}C^{\mu\lambda\nu\kappa}_{\phantom{\lambda\mu\nu\kappa}}
-C^{\mu\lambda\nu\kappa}R_{\lambda\kappa} \ \equiv \ B^{\mu\nu}
\ = \  0\, ,
\label{Z42A}
\end{equation}
where $B^{\mu\nu}$  is the {\it Bach tensor\/} and $D_{\alpha}$ the Riemannian covariant derivative.
%
%

We formally define a quantum field theory of gravity
by a functional integral
\begin{equation}
Z \ = \ \sum_i \int_{\Sigma_i} {\cal D}g_{\mu\nu} \ \! e^{i{\cal A}_{\rm conf}}\, .
\label{Z42Ab}
\end{equation}
\vspace{-0.3mm}
Here  ${\cal D}g_{\mu\nu}$ denotes the functional-integral measure
whose proper treatment involves  the Faddeev--Popov gauge
fixing of the gauge symmetry Diff$\times$Weyl$(\Sigma_i)$
plus ensuing Faddeev--Popov determinant~\cite{Mottola}.
Potential local factors $[-\det g_{\mu \nu}(x)]^{\omega}$
with Misner's ($\omega = -5/2$) or De Witt's ($\omega = (D-4)(D+1)/8$)
are omitted in the measure because they do not contribute to the
Feynman rules. Their effect is to introduce terms $\omega \delta^{(4)}(0) \int d x^4 \log(-g)$
into the action, which by Veltman's rule
are set to zero
in dimensional regularization.
The sum in (\ref{Z42Ab}) is a sum over four-topologies, that is, a sum over topologically distinct manifolds $\Sigma_i$
(analogue to the sum over {\it genus} in string theory or sum over {\it homotopically} inequivalent vacua in the Yang--Mill theory) which
can potentially contain topological phase factors, e.g., Euler number of $\Sigma_i$, cf. Refs.~\cite{Carlip,note}.


It should be remarked that despite the fourth-order nature of the
Bach equation (\ref{Z42A}) indicating the presence of on-shell ghost states~\cite{Riegert,Stelle77},
the recent advances in
non-perturbative~\cite{Ambjorn:12,Biswas,Talaganis,Hamada:09,Egawa:02} and
PT-symmetric~\cite{B-M,B-MII,MIII}
techniques suggest that the would-be ghost states
disappear from the energy eigenspectrum and that CG is stable (i.e.,
non-perturbatively unitary).
Also, the conformal instability typical for the
Euclidean quantum gravity is not presents in CG.
A particularly pleasing aspect of the quadratic-curvature action
(\ref{PA2}) is its
power-counting renormalisability~\cite{noteIIa} and asymptotic freedom
($\beta$-function for $\alpha_c$ is negative)~\cite{Shapiro}.


%

\section{Uncompleting the $R^2$-term}

Here we wish to point out
that the large number of derivatives in the free graviton propagator
implied by (\ref{PA2}) makes fluctuations so violent that the
theory might spontaneously create a new mass term. This phenomenon
is indeed known to happen in number of higher-derivative systems ranging from
biomembranes~\cite{KLMEM} through string theories with extrinsic
curvature~\cite{STK,POL}, to gravity-like theories~\cite{KLMG}. For
instance, in biomembranes and stiff strings the ensuing mass term
can be identified with a tension. We shall now show that an
analogous mechanism spontaneously generates the Starobinsky
action~\cite{Starobinsky}
\begin{equation}
{\cal A}_{\rm St} \ = \
- \frac1{2\kappa^2}\int d^4x\, (-g)^{1/2}
(R - \xi^2 R^2)
\, .
\label{PA5}\\[-1mm]
\end{equation}
Here $\kappa^2  = {8 \pi G_{\rm N}}$ where $G_{\rm N} = 1/m_p^2$ is
Newton's (gravitational) constant and $m_p$ is the Planck mass.
Starobinsky's parameter $\xi$ is related to the inflational scale
and by the Planck satellite data $\xi/\kappa \sim 10^5$ (cf.
Ref.~\cite{Planck}). The minus sign in front of $R^2$-term is a consequence
of the Landau--Lifshitz convention~\cite{Ketov}.

In order to see how the spontaneous generation of (\ref{PA5}) comes
about we first observe that the $R^2$-part of the action (\ref{PA2})
is the global scale-invariant (``${\rm gsi}$'') expression. This is because under
infinitesimal Weyl transformation $g_{\mu\nu} \rightarrow g_{\mu\nu}
+ 2\alpha(x) g_{\mu\nu}$  while $R \rightarrow [1-2\alpha(x)]R - 6
D^2 \alpha(x)$ (the covariant derivative $D_{\mu}$ is with respect
to $g_{\mu\nu}$). Since $g \rightarrow [1 + 8\alpha(x)] g$, the
$R^2$-term part of the action will be scale invariant provided $D^2
\alpha(x) = 0$.
The $R^2$-part of the action can be further decomposed by using the
Hubbard--Stratonovich (HS)
transformation~\cite{Hubbard,Stratonovich}
\begin{eqnarray}
&&\mbox{\hspace{-11mm}}\exp(i{\cal A}_{\rm gsi}) \ \equiv \ \exp\left(\frac{i}{12 \alpha^2_c} \int d^4 x \, (-g)^{1/2} R^2\right)\nonumber \\[2mm]
&&\mbox{\hspace{-11mm}}= \ \int {\cal D} \lambda \ \! \exp\left[-i \!\!\int d^4
x \, (-g)^{1/2} \left(\frac{3 \alpha^2_c}{4} \ \!\lambda^2 + \frac{\lambda
R}{2} \right) \right]\!.
\label{PA2P}
\end{eqnarray}
\vspace{-0.3mm}
%
%
%
%

The essence of the HS transformation is a straightforward manipulation of a Gaussian
integral, which allows to decouple quadratic (or generally quartic) terms in the action
in terms of an auxiliary (bosonic) field variable whose fluctuations can in
principle be described by higher loop diagrams. Due to radiative correction the HS field
can develop in the infrared regime a gradient term which then allows to identify
the HS boson with a genuine dynamical particle. A paradigmatic example of this scenario
is obtained when reducing the BCS superconductivity to its low-energy effective level. There the
HS boson coincides with the disordered field whose dynamics is described via the famous
Ginzburg--Landau equations~\cite{Altland-Simons}.

The HS transformation has currently a well-established place in
solid-state theory~\cite{Altland-Simons,Sachdev} and elementary particle physics~\cite{kl-note,coleman-book}.
It has led to a good  understanding of important collective physical phenomena
such as superconductivity, superfluidity of He$^3$, plasma
and other charge-density waves, pion physics and chiral
symmetry breaking in quark theories~\cite{kl_a}, etc.

Although the auxiliary field $\lambda(x)$ in (\ref{PA2P}) does not have a bare kinetic term,
the local conformal symmetry of ${\cal A}_{\rm conf}$ allows to rescale the metric so that
a kinetic term can easily be generated. For instance, when $g_{\mu\nu} \mapsto |{\lambda}|^{-1} g_{\mu\nu}$ then ${\cal A}_{\rm gsi} $ goes to
\begin{eqnarray}
\int d^4 x \, (-g)^{1/2} \left(-\frac{\lambda R}{2 |\lambda|} + \frac{3 }{4
\lambda^2}\ \!\partial_{\mu} \lambda \partial^{\mu} \lambda   - \frac{3
\alpha^2_c}{4} \right)\!,
\end{eqnarray}
(and other higher-order derivatives of $\lambda$ will come from  the remaining $R^{\mu\nu}R_{\mu\nu}$-term).
Since the $\lambda$-kinetic term depends on the conformal scaling, $\lambda$-kinematics is gauge dependent, implying that $\lambda$ cannot represent a physical field. On the other hand, when the conformal symmetry breaks down
then the $\lambda$-field is trapped in a particular (broken) phase with specific kinetic and potential terms.
This will be shown below.
%

To proceed, we separate the $\lambda$-field  into a background field $\bar\lambda$ corresponding to the VEV of $\lambda$ and fluctuations $\delta\lambda$ which have only nonzero
momenta. Of course, the fluctuations must be included to
make the  theory  completely equivalent  to the original (\ref{Z42Ab}). In the following we employ the standard
effective-action strategy, i.e., neglect  all terms involving $\delta\lambda$, and take the
saddle-point approximation to the remaining integral over $\bar\lambda $.

As will be seen shortly, $\lambda$ spontaneously develops a positive VEV, so that the sign of the $R$-term in (\ref{PA2P})
coincides with the sign of the Einstein term. Since we expect that our theory will eventually induce Einstein's action
(at least at low enough energies) it is convenient
to rescale $\lambda \rightarrow \lambda/\kappa^2$.
%
With the benefit of hindsight we further introduce an {\it arbitrary} mixing angle
$\theta$ and write formally ${\cal A}_{\rm gsi} =
\C^2 {\cal A}_{\rm gsi} -
\S^2 {\cal A}_{\rm gsi}$  where $\C\equiv\cosh \theta$, $\S\equiv\sinh \theta$.
Applying the HS-transformation only to the $(\S^2 {\cal A}_{\rm gsi})$-part we get, after a formal
replacement $\alpha_c^2 \rightarrow -\alpha_c^2/S^2$ in (\ref{PA2P})
\begin{eqnarray}
{\cal A}_{\rm gsi} \ \! &=&\ \!
\frac{C^2}{12\alpha_c^2}\!\int \!d^4x\, (-g)^{1/2}  R^2
\ \! - \ \!\frac{1}{2\kappa^2}\!\int \!d^4x\, (-g)^{1/2}
\lambda R
\nonumber \\[1mm]&&\!\!\!\!\!\!\!\!\hspace{0em} 
+ \  \frac{3\alpha_c^2}{4S^2\kappa^4}\!\int\! d^4x\, (-g)^{1/2}
\lambda^2\, .
\label{@3rdac}\end{eqnarray}
%

Let us now show that the fluctuations of the metric $g_{\mu\nu}$
can achieve the aforementioned scenario. In particular, we find a set of parameters in
the model parameter space  for which $\bar\lambda \equiv \langle \lambda \rangle =1$.
As a result, the long-range behavior of our theory will coincide with
that of  Starobinsky's $f(R)$-model.

\section{Emergence of Starobinsky's model}

We proceed by splitting the  spacetime metric into the flat Minkowski background plus a fluctuation $h_{\mu\nu}$ defined by  $g_{\mu\nu}=\eta_{\mu\nu} +  \alpha_c h_{\mu\nu}$ (realizing that $\alpha_c \sim C \kappa/\xi$), and then expanding  the Lagrangian in (\ref{PA2}) (including the explicit form (\ref{@3rdac}))  to the 2nd order in  $\alpha_c$. Omitting
total derivatives, using the weak-field relations of Appendix~A
%
 %
%
%
and setting $\lambda = \bar{\lambda}$, we end up with the following outcome ($\square \equiv \partial^2$)
%
\begin{eqnarray}
\mbox{\hspace{-1mm}}-{\cal A}_{\rm conf} \ \! &=& \ \!   \frac{1}{16} \int d^4x\ \! h^{\mu \nu} \square^2 h_{\mu \nu}
\ \!  - \ \!  \frac{1}{8} \int d^4x\ \! \partial_{\lambda} h^{\lambda \mu} H_{\mu \nu} \partial_{\rho} h^{\rho \nu}\nonumber\\[1mm]
\ \!  &+& \ \!  \frac{1}{4}\left(\frac{1}{4} -\frac{C^2}{3}\right) \int d^4x\ \! \bar{h} \square^2 \bar{h}\nonumber \\[1mm]
\ \! &+& \ \!  \frac{1}{2\kappa^2} \int d^4x\ \! \left( { - \alpha_c\bar{\lambda} \square \bar{h}}  \ \! - \ \!   \frac{{\alpha}_c^2}{4 }\bar{h} \bar{\lambda} \square \bar{h} \right.\nonumber \\[1mm]
\ \! &-& \ \! \left. \frac{\alpha_c^2}{2}\partial_{\lambda} h^{\lambda \mu} \bar{\lambda} \square^{-1}H_{\mu \nu} \partial_{\rho} h^{\rho \nu} \ \!  + \ \!
\frac{\alpha_c^2}{4} h^{\mu \nu} \bar{\lambda} \square h_{\mu \nu}\right) \nonumber \\[1mm]
\ \! &-& \ \! \frac{3\alpha_c^2}{4S^2\kappa^4} \int d^4x\ \! \bar{\lambda}^2 \nonumber\\[3mm]
&=& \int d^4x\ \! h^{\mu \nu} \mathfrak{A}\square^2 h_{\mu \nu}
\ \!  + \ \!  \int d^4x\ \! \partial_{\lambda} h^{\lambda \mu} \mathfrak{B} H_{\mu \nu} \partial_{\rho} h^{\rho \nu}\nonumber \\[1mm]
\ \!  &+& \ \!   \int d^4x\ \! \bar{h} \mathfrak{C} \square^2 \bar{h} \ \!  + \ \!   \frac{3\alpha_c^2}{4S^2\kappa^4} \int d^4x\ \! \bar{\lambda}^2 \, , \nonumber \\[4mm]
&&\mathfrak{A} = 1/16 + \alpha_c^2 \bar{\lambda} \square^{-1}/(8 \kappa^2),\;\;\;\;\;\mathfrak{B} = -\mathfrak{A}/2,\nonumber \\[3mm]
&&\mathfrak{C} = \frac{1}{4}\left(\frac{1}{4}- \frac{C^2}{3}\right) \ \! - \ \!  \alpha_c^2 \bar{\lambda} \square^{-1}/(8 \kappa^2)\, ,
\label{@Lin}
\end{eqnarray}
%
where $H_{\mu\nu} =  1/2 \ \! \partial_{\mu}\partial_{\nu} - \square \eta_{\mu\nu}$ and $\bar{h} = h^{\mu}_{\mu} - \partial_{\mu} \square^{-1}\partial_{\nu} h^{\mu\nu}$.
A phenomenologically consistent long-range behavior of the
gravitational field is ensured if $\bar\lambda=1$. To see that such
a solution exists at energies low enough, we calculate the one-loop
contribution to the Minkowski effective action. This is obtained by
functionally integrating out the fields $h_{\mu \nu}$ in the
exponential $e^{i{\cal A}_{\rm conf}}$ in which $\lambda$ is
approximated by its VEV, i.e.,  $\bar \lambda$. The result is
$e^{-i\Omega_4 V_{\rm{eff}}}$, where $\Omega_4 $ is the total
four-volume of the universe, and $V_{\rm{eff}}$ is the effective
potential. The form (\ref{@Lin}) is particularly convenient for the
gauge fixings~\cite{Antoniadis,Tseytlin}: $\chi^{\nu} \equiv
\partial_{\mu} h^{\mu\nu} = \zeta^{\nu}(x)$ (coordinate gauge) and
$\chi \equiv  \bar{h} = \zeta(x)$ (conformal gauge). Here
$\zeta^{\nu}(x)$ and $\zeta(x)$ are arbitrary functions of $x$.
Using 't~Hooft's averaging trick~\cite{Hooft:71}:
\begin{eqnarray}
\delta[\chi - \zeta] \ &\rightarrow& \ \int {\mathcal{D}} \zeta \ \!
e^{i\int \zeta {\mathcal{H}} \zeta}
(\det {\mathcal{H}})^{1/2} \delta[\chi - \zeta] \nonumber\\
&=& \ e^{i\int \chi {\mathcal{H}} \chi} (\det {\mathcal{H}})^{1/2}\,
,
\end{eqnarray}
(${\mathcal{H}}$ is an arbitrary symmetric operator) and doing some
straightforward computations we obtain  the zero-genus (fixed topology) contribution to partition function
%
%
\begin{eqnarray}
Z_0 \ &=& \   {\mathcal{N}} (\det  \mathcal{M}_{\rm FP}) (\det H_{\mu \nu}
\det(\square^2)_{\bar{h}})^{1/2}  [\det(-\square^2)_{h_{\mu \nu}}]^{-1/2}\nonumber \\[1mm]
&\times& (\det\mathfrak{C})^{1/2} (\det \mathfrak{A})^{-3} \ \! e^{-i\Omega_4 3\alpha_c^2
\bar{\lambda}^2/(4 S^2 \kappa^4)  }\nonumber \\[1mm]
&=& \mathcal{N}\{[\det(-\square)]^{-1/2}\}^{6}(\det
\mathfrak{C})^{1/2}(\det \mathfrak{A})^{-3} \nonumber \\[1mm]
&\times& e^{-i\Omega_4 3\alpha_c^2
\bar{\lambda}^2/(4
S^2 \kappa^4)  }\, ,
\label{Z0.12}
\end{eqnarray}
%
($(\mathcal{M}_{\rm FP})_{\mu\nu} = -\square \eta_{\mu\nu}  - \partial_{\mu} \partial_{\nu}$ is the
Faddeev--Popov operator for coordinate gauge~\cite{note3}). The factor $\{[\det(-\square)]^{-1/2}\}^{6}$ correctly indicates that that number of propagating modes in the linearized CG is $6$ (cf. Ref.~\cite{Riegert}).
From (\ref{Z0.12}) the one-loop $V_{\rm{eff}}$ reads
\begin{eqnarray}
\mbox{\hspace{-4mm}}V_{\rm{eff}}&=& \frac{i}{2}\!\!\int' \!\! \frac{d^Dk}{(2\pi)^D}\ln\!\left(
k^2 - \frac{6\alpha_c^2 \bar{\lambda}}{\kappa^2 (4\S^2 +1)}\!\right)\nonumber \\[1mm]
&-&\frac{6i}{2} \!\!\int
\!\! \frac{d^Dk}{(2\pi)^D}\ln\!\left(k^2 - \frac{2\alpha_c^2 \bar{\lambda}}{\kappa^2}\!\right)
-
\frac{3\alpha^2}{4S^2
\kappa^4}
\bar\lambda^2.
\label{@.13}
\end{eqnarray}
The prime
indicates a trivial subtraction of the zero-mode. Note that for
(assumed) $\bar{\lambda} > 0$ the ensuing massive pole is physical
only when $\theta \in (-\mbox{arcsinh}(1/4),\infty)$. The integral over $k$ can be evaluated,
e.g.,  with the help of dimensional regularization ($D=4-2\epsilon$)
in  which case it yields
\begin{eqnarray}
V_{\rm{eff}}&=& - \ \frac{9\alpha^4_c\bar{\lambda}^2}{16\pi^2\kappa^4(4S^2 +1)^2}
\left[\ln\frac{6\alpha_c^2\bar\lambda }{ (1+4 S^2) \kappa^2 \varLambda^2} -\frac{3}{2}\right] \nonumber \\[1mm]
&& +\ \frac{3\alpha^4_c\bar{\lambda}^2}{8\pi^2\kappa^4}
\left[\ln\frac{2\alpha_c^2\bar\lambda }{\kappa^2 \varLambda^2} -\frac{3}{2}\right]
\ -\
\frac{3\alpha^2_c}{4S^2
\kappa^4}\bar
\lambda^2\, ,
\label{@22a}\end{eqnarray}
where $\varLambda = \sqrt{4\pi} \mu e^{-\gamma/2} e^{1/2\epsilon}$,
$\mu$ is an arbitrary renormalization scale and $\gamma$ is the
Euler--Mascheroni constant. To obtain a finite result as $\epsilon
\rightarrow 0$ we utilize the $\overline{\mbox{MS}}$ renormalization
scheme. This fixes the counterterm so that
\begin{eqnarray}
V_{\rm{eff}}&=& -
\frac{9\alpha^4_c\bar{\lambda}^2}{16\pi^2\kappa^4(4S^2 + 1)^2}
\left[\ln\frac{6\alpha_c^2\bar\lambda }{ (1+4 S^2) \kappa^2 \mu^2} -\frac{3}{2}\right] \nonumber \\[1mm]
&& +\ \frac{3\alpha^4_c\bar{\lambda}^2}{8\pi^2\kappa^4}
\left[\ln\frac{2\alpha_c^2\bar\lambda }{\kappa^2 \mu^2} -\frac{3}{2}\right] \ -
\
\frac{3\alpha^2_c}{4S^2
\kappa^4}\bar \lambda^2\, , \label{@22abc}\end{eqnarray}
%
%
with $\mu^2$ being the subtraction point.

The saddle point in $\bar\lambda$  corresponding to the VEV is
determined by the vanishing of $V_{\bar\lambda}\equiv\partial V_{\rm
eff}/\partial _{\bar\lambda}$. This
%
%
yields the {\it minimal} $V_{\rm eff}$ for
\begin{eqnarray}
\bar\lambda(S) &=& \exp \left(\frac{3 \alpha_c^2 S^2 \ln \left(\frac{3}{4
S^2+1}\right) + 4 \pi ^2 (4 S^2 + 1)^2}{\alpha ^2 S^2 \left(32 S^4+16
S^2-1\right)}
\right)\nonumber \\[1mm]
&\times& \frac{\kappa^2\mu^2 e }{2 \alpha_c^2}\, .
\label{@Lamda}
\end{eqnarray}
In this case $V_{\rm eff}< 0$ for $S^2 > (\sqrt{6} -2)/8 \approx 0.056$,  irrespective of  actual values  of  $\alpha$ and $\kappa$.
A trivial solution of  $V_{\bar\lambda} = 0$, namely $\bar\lambda(S) = 0$ yields $V_{\rm eff}= 0$ and hence it represents a
local maximum (i.e., unstable solution) for the above range of $S^2$.

Although the full theory described by the action (\ref{@Lin}) is
independent of the mixing angle  $\theta$, the truncation of the
perturbation series after a finite loop order in the fluctuating
$h_{\mu\nu}$-field spoils this independence. The optimal result is
obtained by utilizing the {\it principle} of {\it minimal
sensitivity}~\cite{Stevenson} known from the renormalization-group
calculus. The principle of minimal sensitivity is at the heart of
the $\delta$-perturbation expansion~\cite{BenderIII} and variational
perturbation expansion~\cite{PI,KS22}. There, if the perturbation
theory depends on an {\it unphysical} parameter, say $\theta,$ the best result is
achieved if each order has the weakest possible dependence on the
parameter $\theta$.
Consequently, at the one-loop level the value of $\theta$ is
determined from the vanishing of the derivative of $V_{\rm{eff}}$
with respect to  $S^2$. By setting $V_{S^2}\equiv
\partial V_{\rm eff}/\partial S^2$, we have
%
\begin{eqnarray}
\frac{d V_{\rm eff}}{d S^2} \ = \ \frac{\partial
\bar\lambda(\theta)}{\partial S^2} \ \! V_{\bar{\lambda}} + V_{S^2} \ = \
V_{S^2} \ = \ 0\, .
\label{OP}
\end{eqnarray}
%
This is equivalent to the equation
\begin{eqnarray}
\mbox{\hspace{-2.5mm}}\frac{\left(128 S^6 + 96 S^4 + 36 S^2-1\right)}{S^4
\left(32 S^4+16 S^2-1\right)}\! = \! \frac{12 \alpha^2 \ln
   \left(\frac{4 S^2+1}{3}\right)}{\pi^2 \left(32 S^4+16 S^2 -1\right)},
\end{eqnarray}
which admits two branches of real solutions; either
$S^2 = 0.0259237 - 0.0000197 \alpha^2 + \mathcal{O}(\alpha^4)$ which,
however, does not give a stable $\bar{\lambda}(S)$ (as $V_{\rm eff} > 0$)
or $S$ should have  maximally allowable value within the range of
validity
of our one-loop approximations. This gives the $\bar{\lambda}(S)$-stable
solution $S \sim \xi/\kappa \sim 10^5$.
Consequently, from Eq.~(\ref{@Lamda}) we deduce the one-loop VEV (to order
$\mathcal{O}(1/S^4)$)
\begin{eqnarray}
\mbox{\hspace{-1mm}}\bar\lambda \ = \  \frac{\kappa^2\mu^2 }{2 \alpha_c^2} \ \!
e^{1+{2\pi^2}/{\alpha_c^2 S^2}}\ \sim  \ \frac{\kappa^2\mu^2 }{2 \alpha_c^2}\ \!
e^{1 + {2\pi^2 \kappa^2}/{\alpha_c^2 \xi^2}}\, .
\label{@33b}
\end{eqnarray}
In particular, for any value of  the  dimensionless coupling strength $\alpha_c$,
we can choose the  renormalization mass scale $\mu$, in such a way
that  $\bar\lambda$ has the value $1$, that will  guarantee
phenomenologically correct gravitational forces at long distances.
VEV $\bar\lambda$ is thus the {\it dimensionally transmuted} parameter
of the massless CG. Its role here is completely analogous to the role of the
dimensionally transmuted coupling constant in the Coleman--Weinberg
treatment of the massless scalar electrodynamics~\cite{CWE}. Namely, we have
traded a dimensionless parameter $\alpha_c$ for a  dimensionfull parameter $\bar{\lambda}/\kappa^2$
(which does not exist in the symmetric phase).

By assuming that in the broken phase a cosmologically relevant metric is that of
Friedmann--Lama\^{\i}tre--Robertson--Walker (FRLW),
then, modulo a topological term, the additional condition
\begin{eqnarray}
\int d^4x \ \!  (-g)^{1/2} 3R_{\mu\nu}R^{\mu\nu}  \ = \ \int d^4x \ \!  (-g)^{1/2}  R^2 \, ,
\label{FLRW}
\end{eqnarray}
holds due to a conformal flatness of the FRLW metric~\cite{Birrell}. Combining (\ref{@3rdac}),
(\ref{@33b}), and (\ref{FLRW}),
the low-energy limit of ${\cal A}_{\rm conf}$ in the broken phase reads
\begin{eqnarray}
\mbox{\hspace{-2mm}}{\cal A}_{\rm conf.b.} \ = \ -
\frac1{2\kappa^2}\int d^4x\, (-g)^{1/2}
(R - \xi^2 R^2 - 2\Lambda) \, ,
\label{cosm.const.aa}
\end{eqnarray}
with
\begin{eqnarray}
\xi^2  \ = \ \frac{\kappa^2S^2}{6\alpha_c^2}\, , \;\;\;\;\; \Lambda \ = \ \frac{3 \alpha_c^2}{4S^2\kappa^2}\, .
\label{cosm.const.a}
\end{eqnarray}
We stress that our $\Lambda$ is entirely of a geometric origin (it descends from the CG) and it
enters in (\ref{cosm.const.aa}) with the {\it opposite} sign in comparison with
the usual matter-sector induced (de Sitter) cosmological constant. Note the non-trivial relation between $\xi$ and $\Lambda$, namely $\Lambda = 1/(8\xi^2)$.

\section{Gradient term for $\lambda$}

The local conformal symmetry dictates that
the scalar degree of freedom must decouple from the on-shell spectrum of the
CG~\cite{Tseytlin,Riegert}, whereas in  theories without conformal invariance (but with the
same tensorial content) the scalar field does
appear in spectrum~\cite{Modesto,Riegert,Stelle77}. When the conformal symmetry is broken
the scalar field reappears through a radiatively induced gradient
term of the spurion field $\lambda$.
The explicit form of the kinetic term (namely its overall sign!) can be decided from the momentum-dependent part of
the $\lambda$ self-energy $\Sigma_{\lambda}$. This can be streamlined by considering in (\ref{@Lin}) slowly fluctuating $\lambda$ instead of fixed $\bar{\lambda}$. Since the lowest-order contribution to $\Sigma_{\lambda}$ comes from coupling to $\bar{h}$, the only relevant
substitutions in (\ref{@Lin}) are; $\bar{\lambda} \square \bar{h} \mapsto \lambda \square
\bar{h}$ (which stops to be a total-derivative) and $\bar{h} \bar{\lambda} \square \bar{h} \mapsto
{\lambda} {h}^{\mu\nu} 3 P^{(0)}_{\mu\nu, \alpha\beta}\square {h^{\alpha\beta}}=
\pi_{\mu\nu}(\lambda {h}^{\mu\nu})\square \bar{h}$ ($P^{(0)}_{\mu\nu, \alpha\beta} = \pi_{\mu\nu}\pi_{\alpha\beta}/3$ is the spin-$0$ projection, and $\pi_{\mu\nu} = \eta_{\mu\nu} -\partial_{\mu}\square^{-1} \partial_{\nu}$ is the transverse vector projection). In the leading $\alpha_c$-order, one can neglect $\alpha_c^2 \partial_{\mu} \lambda$
with respect to $\alpha_c \partial_{\mu} \lambda$ and complete
the square in (\ref{@Lin}) as follows
\begin{eqnarray}
&&\mbox{\hspace{-1.5cm}}- \frac{\alpha_c\lambda}{2\kappa^2}\square
\bar{h} \ +  \ \bar{h} \mathfrak{C} \square^2 \bar{h} \nonumber \\[1mm]
&&\mbox{\hspace{-0.3cm}}
 \ \mapsto \ \bar{h} \mathfrak{C} \square^2 \bar{h}  \  -  \
\frac{\alpha^2_c}{16\kappa^4} \square
\lambda (\square^{-2}\mathfrak{C}^{-1} )\square \lambda \nonumber  \\[1mm]
 &&\mbox{\hspace{-0.3cm}}
 \ \approx \ \bar{h} \mathfrak{C} \square^2 \bar{h}  \  +  \ \frac{1}{2\kappa^2\bar{\lambda}} \lambda \square \lambda\, .
 \label{24b}
\end{eqnarray}
The last approximation holds for $ 1 \ll
\alpha_c^2\square^{-1}/(C\kappa)^2 \sim \square^{-1}/\xi^2 \sim 10^{28}
\square^{-1} $, and thus in the large-scale cosmology where only low-frequency modes of scalar fields  (e.g., $\lambda$) are observationally relevant. The square completion procedure employed in (\ref{24b}) changes the
(conformal) gauge fixing condition, albeit the only effect of this modification
is a redefinition of the function $\zeta$.

Because of a minus sign in front of ${\cal A}_{\rm conf}$ in (\ref{@Lin}), the actual kinetic term is
$-\frac{1}{2\kappa^2\bar{\lambda}} \lambda \square \lambda \sim \frac{1}{2\kappa^2\bar{\lambda}} \partial_{\mu} \lambda \partial^{\mu} \lambda$ which is positive. As a result,  $\lambda$ morphs into a genuine (non-ghost, non-tachyonic) propagating scalar mode.

In passing, we note that since $V_{\rm eff}$ in the broken phase is
bounded from below and the kinetic energy is positive (i.e., vacuum decay is prevented), the broken
one-loop linearized CG does not possess ghost states.

%
%
%
This situation is reminiscent of what is seen in a
number of
solid-state systems, including Anderson model, Hubbard model or
superconductors, to name a few.


\section{Cosmological implications}

Recent polarisation data from Planck and WMAP
satellites~\cite{Planck} support inflationary models with
small tensor-to-scalar ratio:
$r  < 0.12$ at 95$\%$ CL. These include, e.g.,
the Starobinsky model (\ref{PA5}), the non-minimally coupled model
($\propto \phi^2 R/2$) with a $V(\phi)\propto \phi^4/4$-potential, and
an inflation model based on a Higgs field~ \cite{Planck}.
In the SM the linear Einstein term determines the long-wavelength
behavior while the $R^2$-term dominates short distances
and drives inflation.
In phenomenological cosmology, the SM
represents metric gravity with a curvature-driven inflation.
In particular, it does not
contain any fundamental scalar field that could be an inflaton, even though a scalar field/inflaton formally appears when transforming the SM to the Einstein frame~\cite{Whitt}.

SM emerges naturally in CG in the weak-field sector of the broken phase where the action ${\mathcal{A}}_{\rm{conf.b.,\lambda}}$ reads
\begin{eqnarray}
&&\mbox{\hspace{-4mm}}
 -
\frac1{2\kappa^2}\int d^4x\, (-g)^{1/2}
\left(\lambda R - \xi^2 R^2 - \frac{(\partial_{\mu} \lambda)^2}{\bar{\lambda}} -
2\Lambda \lambda^2\right) \nonumber \\[2mm]
&&\mbox{\hspace{-4mm}} \stackrel{\lambda \rightarrow \bar{\lambda}
= 1 }{\longrightarrow} -
\frac1{2\kappa^2}\int d^4x\, (-g)^{1/2}
(R - \xi^2 R^2 - 2\Lambda) \, .
\end{eqnarray}
Similarly, as in the usual SM, one can set up for
${\mathcal{A}}_{\rm{conf.b.,\lambda}}$ a dual description
in terms of a non-minimally coupled auxiliary
scalar field $\phi$  with the action~\cite{Duff,Whitt}
\begin{eqnarray}
{\mathcal{A}}_{\phi,J}  &=&  - \frac{1}{\kappa^2}\int d^4 x
(-g)^{1/2}\left(\frac{\lambda + 2\xi \phi}{2}\ \!R  + \frac{\phi^2}{2} \right.\nonumber
\\[2mm]
&& \left.
- \frac{(\partial_{\mu} \lambda)^2}{2\bar{\lambda}} - \Lambda
\lambda^2\right).\label{jordan.fr.1}
\end{eqnarray}
This is a HS-transformed ${\mathcal{A}}_{\rm{conf.b.,\lambda}}$ with $\phi$ being the HS-field.
To analyze (\ref{jordan.fr.1}) we choose to switch from the Jordan frame (\ref{jordan.fr.1}) to the Einstein frame~\cite{Duff,Faraoni,note5a}
where the curvature $R$ enters without a non-minimally coupled fields $\lambda$ and $\phi$. This is obtained via
rescaling: $g_{\mu\nu}\ \mapsto (\lambda + 2\xi {\phi})^{-1}g_{\mu\nu}$,
giving
\begin{eqnarray}
&&\mbox{\hspace{-8mm}}{\mathcal{A}}_{\phi,E}  =  - \frac{1}{\kappa^2}\int d^4 x \ \! (-g)^{1/2}\left[\frac{\tilde{R}}{2} -
\frac{3\xi^2 (\partial_{\mu}\phi)^2 }{(\lambda +2\xi \phi)^2} \right.\nonumber \\[2mm]
&&\mbox{\hspace{-8mm}}  -\frac{3\xi (\partial_{\mu}\phi) (\partial^{\mu}\lambda)}{(\lambda +2\xi \phi)^2}  - \frac{(\partial_{\mu}\lambda)^2  }{2\bar{\lambda}(\lambda +2\xi \phi)} - \frac{3 (\partial_{\mu}\lambda)^2}{4(\lambda +2\xi \phi)^2} \nonumber \\[2mm]
&&\mbox{\hspace{-8mm}} \left. + \ \!\frac{\phi^2}{2(\lambda +2\xi \phi)^2} - \frac{\Lambda \lambda^2  }{(\lambda +2\xi \phi)^2}\right].
\label{einst.fr.1}
\end{eqnarray}
The above metric rescaling is valid only for the metric-signature-preserving transformation, i.e., only when  $(\lambda + 2\xi {\phi})>0$.
The action (\ref{einst.fr.1}) can be brought into a diagonal form if we pass from fields $\{\lambda, \phi\}$ to $\{\lambda, \psi\}$ where the  new field  $\psi$ is obtained via the redefinition
$\phi = [\exp(\sqrt{2/3} |\psi|) -\lambda]/(2\xi)$.
In terms of  $\psi$ the action reads
\begin{eqnarray}
&&\mbox{\hspace{-6mm}}{\mathcal{A}}_{\psi, E} =  - \frac{1}{\kappa^2}\int d^4 x (-g)^{1/2}\left[\frac{\tilde{R}}{2} -
\frac{1}{2}(\partial_{\mu}\psi)^2  + U(\psi,\lambda)  \right. \nonumber \\[2mm]
&&\mbox{\hspace{4mm}}\left.
\  -e^{-\sqrt{2/3}|\psi|}\ \!\frac{(\partial_{\mu}\lambda)^2}{2\bar{\lambda}}\right],
\label{einst.fr.2}
\end{eqnarray}
where $U(\psi,\lambda) = \frac{1}{8\xi^2}\left( 1- 2\lambda e^{-\sqrt{2/3}|\psi|}\right)$,
with $\xi$  from (\ref{cosm.const.a}). The strength of $\lambda$-field oscillations is controlled by the size of a coefficient in front of the $\lambda$-gradient term~\cite{PI}, i.e.,  $e^{-\sqrt{2/3}|\psi|}/\kappa^2$ (more precisely, the local fluctuations square width $\langle (\lambda(x) - \bar{\lambda})\rangle^2 \sim \kappa^2 e^{\sqrt{2/3}|\psi(x)|}$).
At large values of the dimensionless scalar field $\psi$, i.e., at values of the dimensionful field $\tilde{\psi} = \psi/\kappa$ that are large compared to the Planck scale, the gradient coefficient is very small and $\lambda$-field severely fluctuates. Assuming that CG was broken before the onset of inflation, then after a brief period of violent oscillations the $\lambda$-fluctuations are strongly damped~\cite{note4} at $\tilde{\psi} \lesssim 10 m_p$. From then on, the $\lambda$-field settles at its potential minimum at $\bar{\lambda} =1$.
Note that $U(\psi,\bar{\lambda})\leq 1/(8\xi^2) \ll m_p^2$, which is a necessary condition for a successful inflation.  At values of $\tilde{\psi} \sim 10 m_p$, the potential $U(\psi,\bar{\lambda})$ is sufficiently flat to produce the phenomenologically acceptable slow-roll inflation, with the (collective) scalar field $\psi$
playing the role of inflaton.
Using the slow-roll parameters
\begin{eqnarray}
&&\mbox{\hspace{-11mm}}\epsilon \ = \ \frac{1}{2}m_p^2 \left(\frac{\partial_{\psi}U(\psi,\bar{\lambda})}{U(\psi,\bar{\lambda})} \right)^2, \;\;\;\; \eta \ = \ m_p^2 \ \!\frac{\partial^2_{\psi}U(\psi,\bar{\lambda})}{U(\psi,\bar{\lambda})}\, ,
\end{eqnarray}
($\partial_{\psi} \equiv \partial/\partial\psi$) one can write down the tensor-to-scalar ratio $r$ and the spectral index $n_s$ in the slow-role approximation as~\cite{Planck}
\begin{eqnarray}
r \ = \ 16  \epsilon, \;\;\;\; n_s \ = \ 1- 6\epsilon + 2\eta\, .
\end{eqnarray}
In terms of the number $N$ of $e$-folds left to the end of inflation
\begin{eqnarray}
N \ = \ - \kappa^2 \int_{\psi}^{\psi_f} \!\!\! d\psi\ \!  \frac{U(\psi,\bar{\lambda})}{\partial_{\psi}U(\psi,\bar{\lambda})} \ \approx \ \frac{3}{4\bar{\lambda}} e^{\sqrt{2/3}|\psi|} \, ,
\end{eqnarray}
($\psi_{f}$ represents the values of the inflaton at the end of inflation, i.e., when $e^{-\sqrt{2/3}|\psi|}\sim 1$) one gets
\begin{eqnarray}
&&\mbox{\hspace{-10mm}}  n_s \ \approx \ 1- \frac{2}{N},\;\;\;\; r \ \approx \ \frac{12}{N^2}\, ,
\end{eqnarray}
which for $N = 50 \div 60$ (i.e., values relevant for the CMB) is remarkably consistent with Planck data~\cite{Planck}.

While during the inflation, the $\lambda$-field is constant (due to a large coefficient in front of the gradient term) allowing a {\it large-valued} inflaton field  descend slowly from potential plateau,
inflation ends gradually when  $\lambda$ regains its canonical kinetic term, and a {\it small-valued} inflaton field picks up kinetic energy.
%
From  (\ref{einst.fr.2}) the dominant interaction channel at small $|\psi|$  is  $(\partial_{\mu}\lambda)^2 |\psi|$, hence the vacuum energy density stored in the
inflaton field is transferred to the $\lambda$ field via inflaton
decay $\psi \rightarrow \lambda + \lambda$ (reheating), possibly preceded by a non-perturbative stage (preheating).


Note also, that the gravi-cosmological constant $\Lambda$ that was instrumental in setting the inflaton potential in (\ref{einst.fr.2})
has the {\it opposite} sign when compared with
ordinary (matter-sector induced) cosmological constant.
Since the conformal symmetry prohibits the existence of a (scale-full) cosmological
constant, the gravi-cosmological constant must correspond to a scale at which the conformal
symmetry breaks, which in turn determines the cut-off scale of the scalaron.
The magnitude of $\xi$ in the SM is closely linked to the scale of
inflation~\cite{Duff}.
Using the values relevant for the CMB with $50-60$ $e$-foldings, the Planck data~\cite{Planck} require $\xi\sim 10^{-13}$GeV$^{-1}$
or equivalently $\xi/\kappa\sim 10^5$.
Thus from~(\ref{cosm.const.a}) the vacuum energy density is $\rho_{\Lambda} \equiv \Lambda/\kappa^2
\sim 10^{-10}(10^{18} \mbox{GeV})^4 \sim 2\times 10^{100}
\mbox{erg}/{\mbox{cm}}^3$, which corresponds to a zero-point energy density of a
scalaron with an ultraviolet cut-off at $10^{15}-10^{16}$GeV. This is in a range
of the GUT inflationary scale.
For compatibility with an inflationary-induced large structure formation the conformal symmetry should be broken before (or during) inflation~\cite{note3b}.  This can be naturally included in a broader theoretical context of ``conformal inflation'' paradigm, which has been the thrust of much of the recent research~\cite{Linde,Costa,Turok,Vanzo1}.
Let us also notice that the existence of a single scalar field with cutoff at the GUT scale and coupled to broken CG  (e.g., $\lambda$ or GUT Higgs field)  would contribute with a {\it positive} zero-point energy that could substantially reduce or eliminate $\Lambda$.

\section{Conclusions}

To conclude, we have shown that
a spurion-field mediated spontaneous
symmetry breakdown of CG is capable of transforming
a purely metric conformal gravity into an effective scalar-tensor gravity.
%
%
This offers a new paradigm for understanding inflationary
and large-scale cosmology.
In particular, we have shown that the low-energy dynamics in the broken phase is
described by a Starobinsky-type $f(R)$-model, which can be mapped on
a two-field hybrid inflationary model. A dimensional transmutation
ties up together the values of Starobinsky's inflation
parameter $\xi$ and the gravi-cosmological constant $\Lambda$. This  fixes the symmetry-breakdown scale
for CG to be roughly the GUT inflationary scale.
Despite its simplicity, the presented paradigm reproduces not only a phenomenologically acceptable
picture of the large-scale Universe that is compatible with the present
Planck and WMAP data, but it also provides a viable mechanism for the reheating.
Last but not least, the negative gravi-cosmological constant could help to reduce the difference
between theoretically estimated $\rho_{\Lambda}^{(\rm{th})}$ and astronomically observed
$\rho_\Lambda^{(\rm{obs})}$. This would be a particularly powerful scenario when the local conformal
symmetry were a true fundamental symmetry above the inflationary
scale~\cite{note 6b}.

If the original BICEP2 collaboration claimed data supporting inflationary models with a large $r > 0.16$ (i.e, large-field)
were true, then the conventional Starobinsky-type inflationary potential would be excluded.
In turn, this would also invalidate the outlined scenario.
Fortunately, a recent joint analysis of BICEP2/Keck Array and Planck data indicate that the original Planck's conclusions
(on which the cosmological part of this paper is based) are still valid. In particular,
BICEP2/Keck Array and Planck data still support inflationary models with small tensor-to-scalar ratio with
an upper limit $r < 0.12$~\cite{BICEP2 Planck 2015}.

\subsection*{Acknowledgments}

We are grateful to G.~'t~Hooft,
P.~Mannheim, H.C.~Ohanian, N.~Mavromatos, Je-An Gu, Kin-Wang Ng, Misao Sasaki, and L.~Modesto for a number of useful comments.
P.J. acknowledges support from the GA\v{C}R Grant No. GA14-07983S.
F.S. thanks ITP, Freie Universit\"at Berlin for warm hospitality during the early stage of this work.

%

\subsection*{Appendix A}

Here we collect some technical points used in the text. The weak-field expansion of  ${\cal A}_{\rm conf}$ is based on the fluctuating field $h_{\mu\nu}$: $g_{\mu\nu} = \eta_{\mu\nu} + \alpha_c h_{\mu\nu}$. This gives ${R}_{\a\b\c\d}  \ = \  \frac{\alpha_c}{2}
           \left[ \partial _\b \partial_\c h_{\a\d} + \partial _\a
            \partial_\d h_{\b\c} - (\a\leftrightarrow \b)\right]$ and to the order $\alpha_c^2$ results in
\begin{eqnarray}
&&\sqrt{-g} \ = \  1 \ + \  \frac{\alpha_c}{2} \ \! h^{\mu}_{\mu} \ + \  \frac{\alpha_c^2}{8}\left(h^{\mu}_{\mu}h^{\nu}_{\nu} - 2 h_{\mu}^{\nu}h_{\nu}^{\mu} \right) ,
\nonumber \\[1mm]
&&\sqrt{-g} {R}_{\a\d}^2\! \ = \ \! \frac{\alpha_c^2}{4}
        (\partial _\a \partial _\c h_{\d}^{\c}\! + \partial _\d \partial _\c
         h_{\a}^{\c}\! - \partial _\a \partial _\d h\! - \partial ^2 h_{\a\d})^2\, , \nonumber\\[1mm]
&&\sqrt{-g}{R}^2  \ = \  \alpha_c^2(\partial ^2 h - \partial _\a \partial _\b
       h^{\a\b})^2   \, .
\label{lingr5.5c}
\end{eqnarray}
%
%
With this the weak-field expansion of the Weyl action (\ref{PA1})
reads (modulo total derivatives)~\cite{Tseytlin}
\begin{eqnarray}
{\cal A}_{\rm conf} \ &=& \ -\frac{1}{8\alpha_c^2}\int d^4x\, (-g)^{1/2}C_{\lambda\mu\nu\kappa}
C^{\lambda\mu\nu\kappa}\nonumber \\
&=& -\frac{1}{16}\int d^4x\,  h^{\mu\nu} P^{(2)}_{\mu\nu, \alpha\beta}\square^2  h^{\alpha\beta} \nonumber \\
&=& -\frac{1}{16}\int d^4x\, (\square \bar{h}^{\alpha\beta, \bot})^2\, .
\label{A1}
\end{eqnarray}
Here, $P^{(2)}_{\mu\nu, \alpha\beta}= \pi_{\mu(\alpha}\pi_{\beta)\nu} -\frac{1}{3} \pi_{\mu\nu}\pi_{\alpha\beta}$ is the spin-$2$ projection, and $\pi_{\mu\nu} = \eta_{\mu\nu} -\partial_{\mu}\square^{-1} \partial_{\nu}$ is the transverse vector projection. We also used $\bar{h}^{\alpha\beta, \bot}$ which is defined via two tensor decompositions: a) $h_{\mu\nu} = \bar{h}_{\mu\nu} + \frac{1}{4}\eta_{\mu\nu}\varphi $ ($\varphi \equiv h^{\alpha}_{\alpha}$ so that $\bar{h}^{\alpha}_{\alpha} = 0$) and b) $\bar{h}_{\mu\nu} = \bar{h}_{\mu\nu}^{\bot} + \partial_{\mu}\eta^{\bot}_{\nu} + \partial_{\nu}\eta^{\bot}_{\mu} + \partial_{\mu}\partial_{\nu} \sigma - \frac{1}{4}\eta_{\mu\nu}\partial^2 \sigma$ (with $\partial^{\mu }\bar{h}_{\mu\nu}^{\bot} = 0$ and $\partial^{\mu}\eta^{\bot}_{\mu}=0 $) which serve to identify irreducible degrees of freedom.  Using the conformal gauge~\cite{Antoniadis,Tseytlin} $\varphi = 0$ and the coordinate gauge $\partial_{\mu}h^{\mu\nu} = 0$ (with the associated Faddeev--Popov operator $(\mathcal{M}_{\rm FP})_{\mu\nu} = -\square \eta_{\mu\nu}  - \partial_{\mu} \partial_{\nu}$) the functional measure reads
\begin{eqnarray}
{\mathcal{D}}h_{\mu\nu} \ &=& \  {\mathcal{D}} \bar{h}_{\mu\nu}^{\bot}{\mathcal{D}} \eta^{\bot}_{\mu} {\mathcal{D}} \sigma {\mathcal{D}} \varphi \det(-\square)_{\sigma} [\det(-\eta_{\mu\nu}\square)_{\eta^{\bot}} ]^{1/2}\nonumber \\
&\mapsto& {\mathcal{D}} \bar{h}_{\mu\nu}^{\bot} [\det(-\square)_{\sigma} \det(-\eta_{\mu\nu}\square)_{\eta^{\bot}} ]^{1/2}\, .
\label{A2}
\end{eqnarray}
For one-loop effective action calculations we used more convenient approach in which the weak-field action ${\cal A}_{\rm conf}$ is written in terms of
{\it unconstrained} variable $h_{\mu\nu}$  as
\begin{eqnarray}
\!\!\! -\frac{1}{16}\!\int \!\!d^4x \left[ h^{\mu\nu} \square^2 h_{\mu\nu} - \partial_{\lambda} h^{\lambda\mu} H_{\mu\nu} \partial_{\rho} h^{\rho \nu} - \frac{1}{6} \bar{h} \square^2 \bar{h}\right]\!.
\end{eqnarray}
Here $H_{\mu\nu} =  1/2 \ \! \partial_{\mu}\partial_{\nu} - \square \eta_{\mu\nu}$ and $\bar{h} = h^{\mu}_{\mu} - \partial_{\mu} \square^{-1}\partial_{\nu} h^{\mu\nu}$. To obtain the diagonal kinetic operator one has to cancel the second and third therm by fixing the gauges: $\chi^{\nu} \equiv
\partial_{\mu} h^{\mu\nu} = \zeta^{\nu}(x)$ (coordinate gauge) and
$\chi \equiv  \bar{h} = \zeta(x)$ (conformal gauge). As before, the Faddeev--Popov operator for coordinate gauge is $(\mathcal{M}_{\rm FP})_{\mu\nu} = -\square \eta_{\mu\nu}  - \partial_{\mu} \partial_{\nu}$ while for conformal gauge $(\mathcal{N}_{\rm FP}) = (D-1)\delta^{(D)}(x-y)$.
In this case the functional-integral measure is:
\begin{eqnarray}
{\mathcal{D}}h_{\mu\nu} \ \mapsto \ {\mathcal{D}}h_{\mu\nu} \delta[\chi - \zeta]\delta[\chi^{\nu} - \zeta^{\nu}] \det(\mathcal{M}_{\rm FP})\, ,
\end{eqnarray}
($\zeta$ and $\zeta^{\mu}$ are arbitrary functions of $x$).
With the help of 't-Hooft's averaging trick the corresponding
partition function coincides with that obtained from (\ref{A1})-(\ref{A2}).

%

\begin{thebibliography}{99}
\bibitem{note2} Non-perturbative techniques have been long indicating that no-ghosts should be present in the CG~\cite{toumboulis,kaku}.
\bibitem{noteIIa} There is still a possibility that CG could be (non-perturabatively) non-renormalisable owing to the conformal anomaly, see~\cite{duffII}.
\bibitem{note3} For conformal gauge the analogous operator is merely an unimportant c-numbered function $(D-1)\delta^{(D)}(x-y)$.
\bibitem{note4} By ``small'' is meant that the relative fluctuations $\sqrt{\langle ({\lambda} - \bar{\lambda})^2\rangle/\bar{\lambda}^2 }
\lesssim 10^{-17}$, i.e., smaller than the GUT scale.
\bibitem{note5a} The issue of (non-)equivalence between Jordan and Einstein frames
is a quite controversial topic in current cosmology and astrophysics.
Comparison of both methods with their respective pros and cons
can be found, e.g., in Refs.~\cite{Duff,Faraoni,Cho}.
\bibitem{note3b} Of course, in order to pinpoint the exact sequence of events a further renormalization group analysis is needed.
\bibitem{note 6b}  If confirmed by subsequent non-perturbative analysis that CG is nevertheless non-unitary, the presented scenario would remain valid with the proviso that CG should be viewed as an effective deeper level dynamics. One can then argue that the true fundamental theory (i.e., the ultraviolet completion of CG) could be Berkowits--Witten twistor-string theory~\cite{witten} or $N=4$ Conformal Supergravity, which both harbor CG  in their low-energy limits (and do not have any pending unitarity issues).
\end{thebibliography}
%

\end{document}